\documentstyle[12pt,psfig]{article}

\begin{document}
\vskip 1.5cm
\centerline{\large\bf Hadron yields in Au+Au/Pb+Pb at RHIC and LHC}
\medskip
\centerline{\large\bf from thermalized minijets}
\medskip
\bigskip
\bigskip
\bigskip
\centerline{\bf N. Hammon, H. St\"ocker, W. Greiner}
\medskip
\centerline{Institut F\"ur Theoretische Physik}
\centerline{Robert-Mayer Str. 10}
\centerline{Johann Wolfgang Goethe-Universit\"at}
\centerline{60054 Frankfurt am Main, Germany}
\bigskip
\centerline{\bf A. Dumitru}
\medskip
\centerline{Physics Department, Yale University}  
\centerline{P.O.\ Box 208124, New Haven, CT 06520, USA}
\bigskip
\medskip
\medskip
\bigskip
\bigskip
\centerline{Abstract}
\bigskip
{\small 
We calculate the yields of a variety of hadrons for RHIC and LHC 
energies assuming thermodynamical equilibration of the produced
minijets, and using as input results from pQCD for the energy densities
at midrapidity.
In the calculation of the production of partons and of 
transverse energy one has to account for nuclear shadowing. By using 
two parametrizations for the gluon shadowing one derives energy densities
differing strongly in magnitude.
In this publication we link those perturbatively calculated energy densities of
partons via entropy conservation in an ideal fluid to the hadron multiplicities
at chemical freeze-out.}
\newpage
\section{Introduction}
Particle production in high-energy heavy-ion reactions at the BNL-RHIC
and CERN-LHC colliders will soon provide interesting insight into nuclear
modifications of semi-hard processes
\cite{Mueller}. This is because
pQCD processes involving gluons in the initial state will dominate the
inelastic AA cross-section at collider energies.
In particular, they might lead
to a better understanding of the gluon
distribution in large nuclei, which is not accessible in DIS. 

In \cite{acapella} the effect of nuclear
shadowing of the parton distribution functions on the charged particle multiplicity 
at midrapidity
has been investigated assuming no rescattering between the produced minijets
and the hadrons they fragment into. Here, we will take the opposite point of
view and assume maximal rescattering, i.e.\ local thermal and chemical
equilibrium of the minijets. We compute final-state hadron multiplicities
of various hadron species under the assumption of entropy conservation.

\begin{figure}
\centerline{\psfig{figure=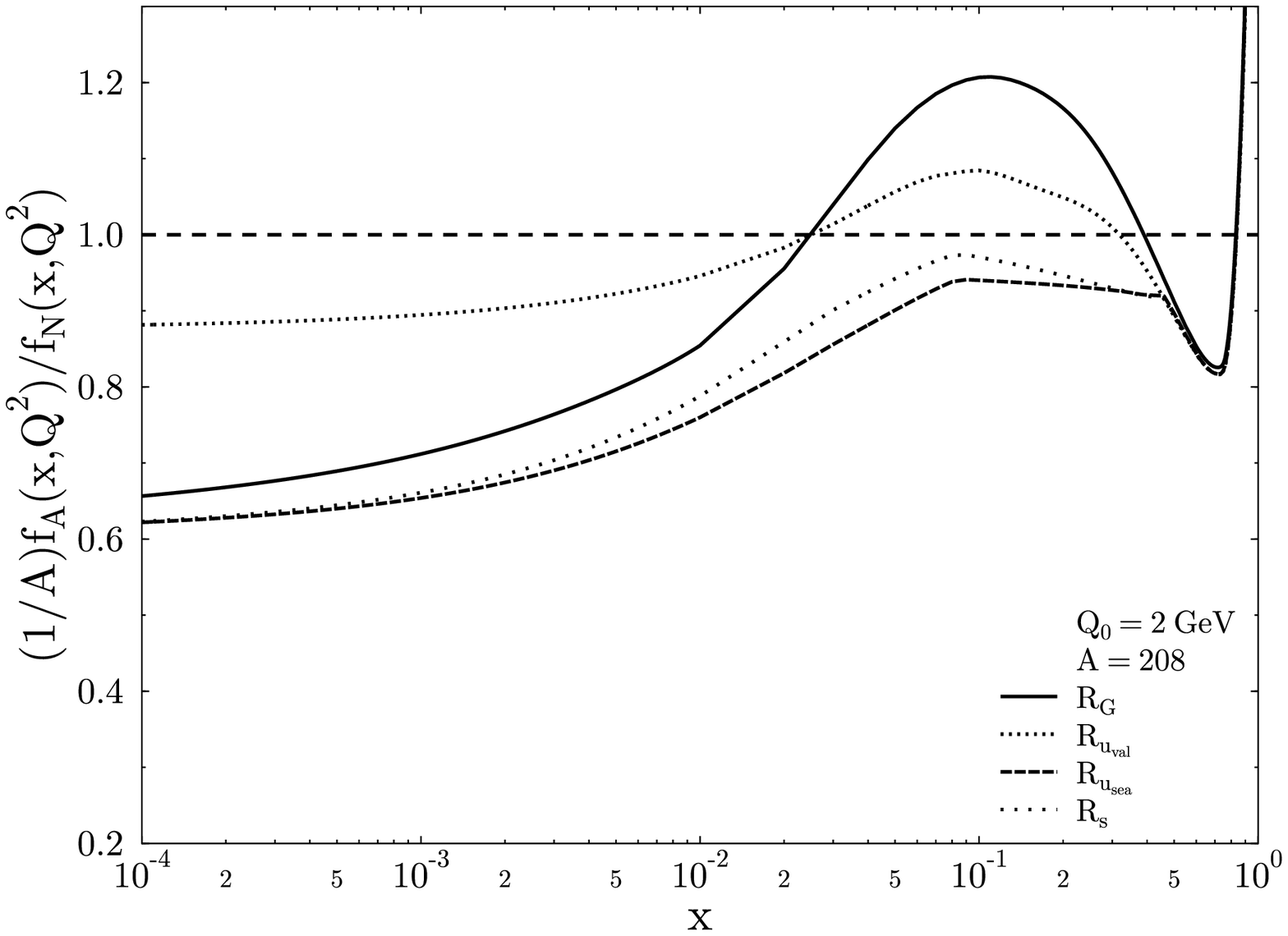,width=10cm}\hspace{-2.8cm}
\psfig{figure=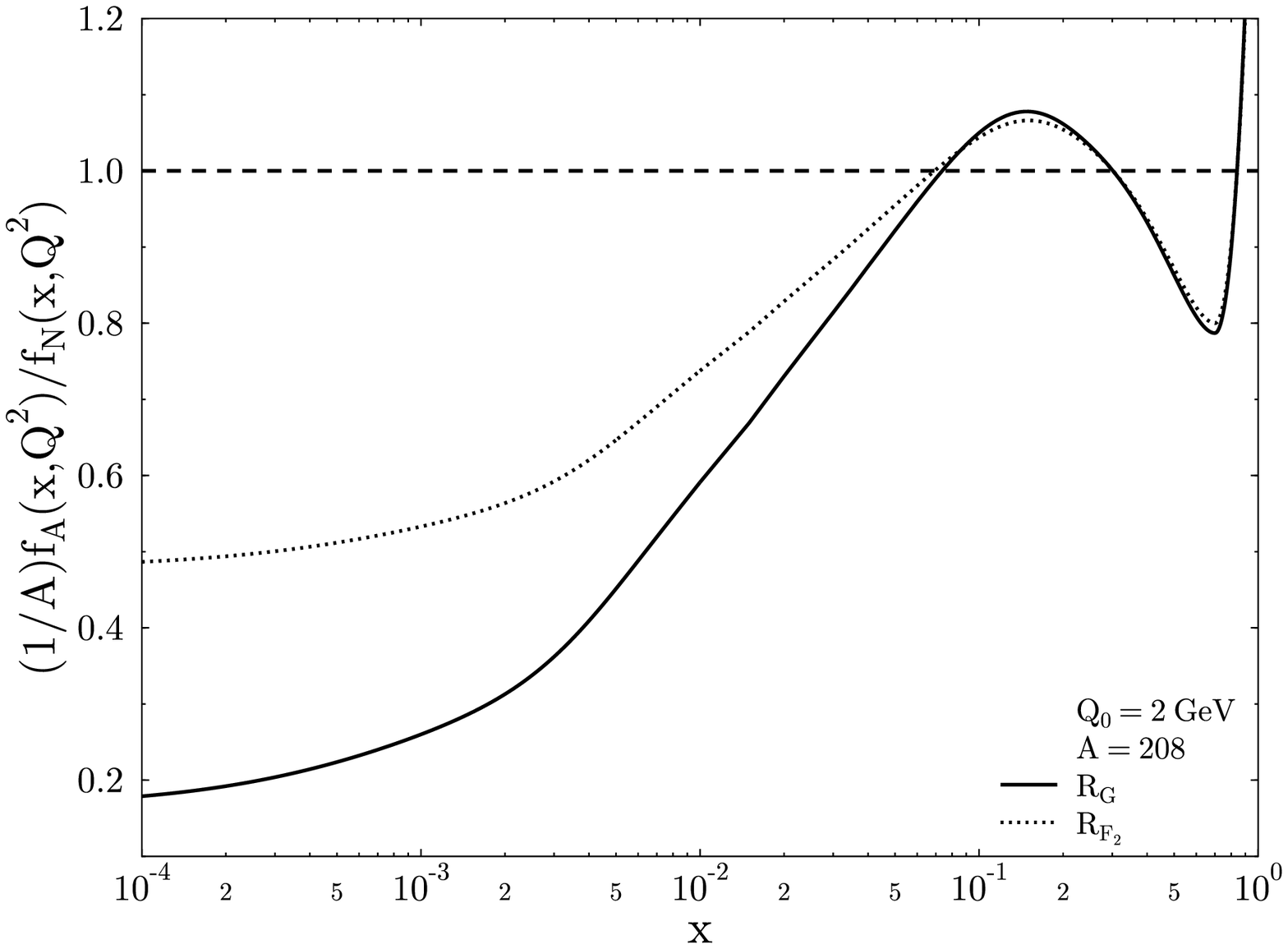,width=10cm}}
\vspace*{-1.5cm}
\caption{\it Shadowed parton distributions as parametrized
in \protect\cite{eskola3} and in \protect\cite{ham99}.}
\label{shad-param}
\end{figure}

In \cite{ham99} we calculated the initial conditions at RHIC and LHC
by means of pQCD above the semihard scale $p_T=2$ GeV to derive the 
number and energy densities of partons at midrapidity. In that calculation we explicitely 
included the shadowing effect on the parton distribution functions entering the
formulas for the production of flavor $f=g, q, \bar q$ in the minijet approach.
We employed two different parametrizations for the shadowing effect accounting
for weak and strong gluon shadowing, respectively, shown in figure \ref{shad-param}. 
A direct consequence of the 
shadowing effect is the decrease in the production of partons of given
momentum $p_T$, i.e.~a decrease of transverse energy production at midrapidity
(for a more detailed description of the calculation, the variables, and the hard partonic
subprocesses see \cite{ham99, eskola1, eichten, kunszt}).\\
We calculate the first $E_T$ moment with and without shadowed pdf's
and with a cut-off function $\epsilon (y)$ ensuring that we only count scatterings
into the central rapidity region ($\left| y\right| \leq 0.5$):
\begin{eqnarray}
&& \sigma^{f}_{hard} \left < E_T\right >_{hard} = \int dE_T \frac{d\sigma ^f}{dE_T} 
\left < E_T\right >
= \int dp_{T}^{2}~ dy_1~ dy_2~ \sum_{ij,kl} ~x_1 f_{i}(x_1,Q^2)~ \nonumber \\
& & x_2 f_{j} (x_2,Q^2) 
\left [ \delta_{fk} \frac{d\hat \sigma^{ij\rightarrow kl}}{d\hat t}
(\hat t, \hat u)+
\delta_{fl} \frac{d\hat \sigma^{ij\rightarrow kl}}{d\hat t}(\hat u, \hat t)
\right ]
\frac{p_T\epsilon (y)}{1+\delta_{kl}}.
\end{eqnarray}
The results for the production of transverse energy by the semihard partons
at RHIC for unshadowed, weakly shadowed, and strongly shadowed gluons 
are shown in table \ref{table1} (for the results from the different subprocesses
see \cite{ham99}).
\begin{table}
\begin{tabular}{|c||c|c|c|} \hline
& no shad. & weak shad. & strong shad. \\ \hline \hline
$\sigma^{g}\left< E_T\right>$ & 26.74 & 27.25 & 17.80 \\ \hline 
$\sigma^{q}\left< E_T\right>$ &  3.20 & 2.87 & 1.00 \\ \hline
$\sigma^{\bar{q}}\left< E_T\right>$ &  1.89 & 1.73 & 0.59 \\ \hline
\end{tabular}
\caption{\it First \protect$E_T$-Moment in {\rm mb GeV} at midrapidity 
(\protect$\left| y \right| \leq 0.5$) for parton production at RHIC for the different 
shadowing parametrizations and no shadowing, respectively.}
\label{table1}
\end{table}   
By dividing the transverse energy $\bar{E}_{T}^{f,AA}(b=0)=T_{AA}(0)
\sigma^{f}_{hard} \left < E_T\right >_{hard}$ of parton species $f$
by the initial volume 
\begin{equation}
V_i = \pi R_{A}^2 \Delta y /p_0,~~R_{A}= A^{1/3} \times 1.1~{\rm fm}
\end{equation}
we found the energy densities for the three different cases as shown in 
table \ref{table2}.
\begin{table}
\begin{tabular}{|c||c|c|c|} \hline
& no shad. & weak shad. & strong shad. \\ \hline \hline
$\varepsilon _g$ & 60.0 & 61.2 &  40.0\\ \hline
$\varepsilon _q$ &  7.2 & 6.4 & 2.3 \\ \hline
$\varepsilon _{\bar q}$ & 4.3 & 3.9 & 1.3 \\ \hline
\end{tabular}
\caption{\it Energy densities for unshadowed, weakly and strongly shadowed gluons in
\protect${\rm GeV/fm}^{3}$ for \protect$Au+Au$ at RHIC.} 
\label{table2}
\end{table}
One sees that the results for the energy density with the weak shadowing are even slightly 
increased due to the onset of $R_G(x,Q^2)$.
We also included the contributions of the soft processes as discussed in
\cite{eskola2}. Motivated by the factorization in QCD one can assume that 
the production of transverse energy in AA collisions can be split up into a 
hard and a soft contribution as
\begin{equation}
\bar{E}_T (b) = T_{AA}(b)\left[ \sigma_{hard}^{pp}\left<E_T\right>_{hard}^{pp}+
\sigma_{s}^{pp}\left<E_T\right>_{s}^{pp} \right].
\end{equation}
With an energy independent value of $\sigma_{s}^{pp}=32$ mb one derives
\cite{eskola2} $\sigma_{s}^{pp}\left<E_T\right>_{s}^{pp}=15$ mb GeV. 
With $T_{AuAu}=29/$mb one can derive the soft 
contribution (i.e.~the one for $p_T\leq 2$ GeV) to the energy density for RHIC
as $\varepsilon _{soft} = 33.7~{\rm GeV/fm}^{3}$.
If one assumes the soft contribution to be independent of the shadowing phenomenon
one can derive the change in relative weight of hard to soft processes due to the shadowing 
of the hard contributions. For the total first $E_T$ moment of gluons, quarks, and antiquarks
one finds that 
$(\sigma^{g}_{hard} + \sigma^{q}_{hard} + \sigma^{\bar q}_{hard})\left<E_T\right>_{hard}$
equals $31.8$ mbGeV for no, $19.4$ mbGeV for strong, and $31.9$ mbGeV for weak shadowing.\\
When comparing to the soft contributions one finds with
$\sigma_{hard}\left<E_T\right>_h = \left(\sigma^g + \sigma^q +
\sigma^{\bar q}\right)_{hard}\left<E_T\right>_{hard}$
that the ratio of soft to hard contribution 
\begin{equation}
R_{sh}=\frac{\sigma_{soft}\left<E_T\right>_s}{\sigma_{hard}\left<E_T\right>_h}
\end{equation}
is $R_{sh}=0.47$ for no, $R_{sh}=0.77$ for strong, and $R_{sh}=0.47$ for weak shadowing, 
respectively.
This implies that at RHIC the soft component could 
even dominate if it were energy independent and also unaffected by the 
shadowing effect.\\

The same analysis was done for LHC and we found
for the first $E_T$ moments in the three cases the results depicted in
table \ref{table3}. For LHC we only calculated the contribution of the gluons that 
strongly dominate all partonic processes due to the large distribution function at
small momentum fractions. 
\begin{table}
\begin{tabular}{|c||c|c|c|} \hline
& no shad. & weak shad. & strong shad. \\ \hline \hline
$\sigma^{g}\left< E_T\right>$ & 513.01 & 286.87 & 60.39 \\ \hline
\end{tabular}
\caption{\it First \protect$E_T$-Moment in {\rm mb GeV} at midrapidity 
\protect$\left| y \right| \leq 0.5$
for gluon production at LHC for the different shadowing 
parametrizations and no shadowing, respectively.}
\label{table3}
\end{table}
The results for the energy densities for no, strong, and weak gluon
shadowing, then are 
$\varepsilon _g = 1229.7~{\rm GeV/fm}^{3}$, $\varepsilon _g = 144.8~{\rm GeV/fm}^{3}$,
and $\varepsilon _g = 678.6~{\rm GeV/fm}^{3}$.  
With $T_{PbPb}=32/$mb one can derive the energy density from the soft part
and has $\varepsilon _{soft} = 35.8~{\rm GeV/fm}^{3}$
which is slightly larger than at RHIC due to the larger nuclear overlap function,
i.e.~the larger number of effective scatterings in the Glauber picture leading to 
the transverse energy production.
For the various shadowing parametrizations one finds for the total first
$E_T$ moment 
$(\sigma^{g}_{hard} + \sigma^{q}_{hard} + \sigma^{\bar q}_{hard})\left<E_T\right>_{hard}$
is $513$ mbGeV for no, $60$ mbGeV for strong, and $287$ mbGeV for weak shadowing.
The relative weight 
\begin{equation}
R_{sh}=\frac{\sigma_{soft}\left<E_T\right>_s}{\sigma_{hard}\left<E_T\right>_h}
\end{equation}
between soft and hard contributions therefore changes and becomes 
$R_{sh}=0.029$ for no, $R_{sh}=0.25$ for strong, and $R_{sh}=0.052$ for weak shadowing, 
respectively.\\
Therefore the soft contributions gain much more weight in this naive picture 
due to the strong effect of shadowing on the small-$x$ gluons
(for further details of the calculation,
the shadowing parametrizations, etc.~see \cite{ham99}).
\section{Hadron Multiplicities at Chemical Freeze-Out}
Having calculated the energy densities at midrapidity for RHIC and LHC in pQCD
we connect $\varepsilon _i$ to the number of hadrons at midrapidity by assuming an 
ideal fluid that is characterized by entropy conservation from the quark-gluon plasma to 
the hadron gas, $dS_i/dy = dS_f/dy$ \cite{Bjorken}.
We relate the energy density of the quark-gluon plasma to its
entropy density via the bag model equation of state \cite{cho}.
We account for $u$, $d$, $s$ quarks (with masses $m_u=m_d=0$, $m_s=150$ MeV),
the antiquarks, and gluons.

The total produced entropy $dS_i/dy$ is obtained from the entropy density
at midrapidity as $dS_i/dy = V_i s_i$ with the initial volume of the central region
$V_i =\pi R_{A}^{2}\tau_i$,
which is numerically $V_i=12.9~{\rm fm}^3$ for $Au+Au$ and $V_i=13.4~{\rm fm}^3$ for 
$Pb+Pb$ at $b=0$ with $\tau =0.1~{\rm fm}/c$. Since we assumed an ideal fluid the 
total entropy is 
conserved throughout the expansion until freeze-out which is chosen here to 
happen at a temperature $T_{FO}=160$ MeV. For simplicity we furthermore assume
vanishing chemical potentials in the central rapidity region, i.e.~that all conserved 
currents are identically zero. If this were not true one would have to multiply by factors
${\rm exp}(\mu_i/T)$ (in Boltzmann approximation).

The entropy density of the hadronic fluid is calculated assuming an ideal gas
composed of all hadrons up to a rest-mass of 2 GeV. Their respective
occupation numbers are given by Fermi-Dirac or Bose-Einstein distribution
functions, respectively. Thus, $T_{FO}$ and $dS_f/dy=dS_i/dy$ determine the
multiplicity of each hadron species uniquely \cite{pbm, adrian}.
Feeding from post freeze-out decays of heavier resonances is also taken
into account.
\section{Results}
With the model outlined above and the energy densities derived
above we calculated the number of a variety of hadrons
at midrapidity. We also include the multiplicities due to the soft contributions
and quote the initial temperatures for a QGP of three flavors.
For LHC we derived the yields shown in table \ref{table4} and for RHIC the ones in
table \ref{table5}.
\begin{table}
\begin{tabular}{|c||c|c|c|c|c|c|c|c|} \hline
{\bf LHC} & $\pi^+ + \pi^-$ & K & $\phi$ & p & $\Lambda$ & $\Xi$ & $\Omega$ & $T_i$ \\ 
\hline \hline
no shad. & 2680 & 478 & 32.1 & 91 & 58 & 9.3 & 1.4 & 881 MeV\\ \hline
weak shad. & 1720 & 306 & 20.6 & 58.3 & 37.2 & 5.9 & 0.9 & 760 MeV \\ \hline
strong shad. & 538 & 95.9 & 6.5 & 18.3 & 11.6 & 1.9 & 0.3 & 516 MeV\\ \hline
soft contrib. & 187 & 33 & 2 & 6 & 4 & 0.7 & 0.1 & \\ \hline
\end{tabular}
\caption{\it Hadron yields at freeze-out with initial conditions from pQCD for LHC.
The initial temperatures are for a three flavor quark-gluon plasma with two
massless quarks and \protect$m_s=150$ {\rm MeV}. Since the soft processes do not 
significantly contribute to the temperature we did not calculate \protect$T_i$
for the soft partons.}
\label{table4}
\end{table}
For the latter one, one can clearly see
that there is no change in the hadron yield for weak shadowing and the unshadowed case, 
respectively, due to the almost identical energy density serving as an input for the
calculation.  
\begin{table}
\begin{tabular}{|c||c|c|c|c|c|c|c|c|} \hline
{\bf RHIC} & $\pi^+ + \pi^-$ & K & $\phi$ & p & $\Lambda$ & $\Xi$ & $\Omega$ & $T_i$ \\ 
\hline \hline
no shad. & 316 & 56.3 & 3.8 & 10.7 & 6.8 & 1.1 & 0.2& 433 MeV\\ \hline
weak shad. & 316 & 56.3 & 3.8 & 10.7 & 6.8 & 1.1 & 0.2 & 433 MeV \\ \hline 
strong shad. & 217 & 38.7 & 2.6 & 7.4 & 4.7 & 0.8 & 0.1 & 383 MeV\\ \hline
soft contrib. & 179 & 32 & 2 & 6 & 4 & 0.6 & 0.1 & \\ \hline
\end{tabular}
\caption{\it Hadron yields at freeze-out with initial conditions from pQCD for RHIC.}
\label{table5}
\end{table}
\section{Conclusions and outlook}
We computed the rapidity densities
of a variety of hadrons 
based on the assumption of entropy conservation of an ideal fluid. The
entropy densities were derived from the energy densities which in turn were 
caluclated by means of pQCD \cite{ham99}. We find that in the limit that the minijets
equilibrate locally the effect of shadowing on the 
hadron yield is not as large as on the pure partonic degrees of freedom.
This can be seen e.g.~in the ratio of energy densities between unshadowed
and strongly shadowed gluons at LHC, which is about a factor of 9, while the ratio of
the hadron yields is only a factor of 5. Since vanishing net baryon and
strangeness densities were assumed, 
the relative depletion of shadowed to unshadowed gluon distribution
is independent of the particle species. For RHIC, even without shadowing
we only get about 300 pions since the perturbative calculation, entering
via the energy densities, was performed with the cut-off $p_T =2$ GeV. Therefore,
the soft contribution constitutes a significant part of the transverse energy 
and therefore of the particle multiplicities \cite{adrian}.
E.g.~in the UrQMD model the total pion yield 
at midrapidity is $\approx 1100$ \cite{bleicher99} for RHIC. Therefore with an energy density of 
$\varepsilon _s = 33.7~{\rm GeV/fm}^3$, which was in extracted in \cite{eskola2}
from CERN data,
and assumed to be energy independent, we get 316+179 pions and therefore get a significantly
smaller multiplicity compared to calculations in the hadronic cascade model UrQMD. The 
key to this discrepancies could stem from the assumed energy independence of the
soft contribution. For a high energy process one can write the interaction as a 
current-current interaction as
\begin{equation}
\sigma (p_T) = H^0 \otimes f^2 \otimes f^2 + \left(\frac{1}{p_T}\right)H^1 \otimes f^3 \otimes f^2
+{\cal O} \left(\frac{1}{p_{T}^{2}}\right)
\end{equation}
where the $f^n$ are non-perturbative IR dominated matrixelement of (lowest) 
twist $n$ and $H$ depicts the hard part.
One should interpret the soft processes at least in part to stem from higher twist 
effects. For an unpolarized process the next-to-leading twist is $\tau$-4. In general 
one can relate the matrix elements of higher twist to the well known
parton distribution functions of $\tau$-2. In \cite{guo} the twist-4 
correlation functions in a nucleus were parametrized as
\begin{equation}
T_{q,g} = \lambda ^2 A^{1/3} f_{i/A}(x,Q^2,A)=
\lambda ^2 A^{1/3}R_{q,g}(x,Q^2,A) f_{i/p}(x,Q^2)
\label{eq-guo}
\end{equation}
where $\lambda ^2 \sim 0.05-0.1 {\rm GeV}^2$. This immediately implies that the soft processes
are not independent of the energy since $x\sim 1/\sqrt s$ and therefore should also 
increase from CERN-SPS to BNL-RHIC/CERN-LHC. Equation (\ref{eq-guo}) in turn implies that 
the soft processes also are affected by shadowing; and since the DGLAP evolution between
$Q\sim1$ GeV and $Q\sim2$ GeV is rather slow, one should expect no change in the 
relative weight between hard and soft processes. As a result one has two competing effects:
as the c.m.~energy increases the soft production of transverse energy increases due to 
the decrease in the momentum fraction $x$ and eq. (\ref{eq-guo}). On the other hand
the shadowing ratio $R_G(x,Q^2)$ suppresses those contributions as $x$ decreases. 
This topic requires a more detailed analysis elsewhere.\\
The interesting feature at RHIC is the dependence of the hadron yields on the onset of 
gluon shadowing which differs substantially in the two approaches we compared. 
Since for the 
weaker gluon shadowing given in \cite{eskola3} the energy density for RHIC
is more or less the same as in the unshadowed case we see the same results for 
the particle numbers in the two cases. However, for the strong gluon shadowing
we see a depletion to about $70 \%$ of the unshadowed value.
The effect of shadowing on the hadron multiplicities at central rapidity is found 
to be weaker than on the newly produced partons.\\
\centerline{\bf Acknowledgements}
This work was supported by BMBF, DFG, and GSI. A.D.~thanks the Yale Relativistic 
Heavy Ion Group for kind hospitality and support from grant 
no.~DE-FG02-91ER-40609.

\end{document}